\documentclass[twocolumn,aps,prl,floatfix,showpacs]{revtex4}
\usepackage{graphicx}
\usepackage{bm}
\usepackage{amsfonts}
\usepackage{amsmath}
\usepackage{dcolumn}

\begin{document}

\title{Athermal All-Optical Femtosecond Magnetization Reversal in GdFeCo}

\author{J. Hohlfeld$^1$}
\email{julius.hohlfeld@seagate.com}

\author{C. D. Stanciu$^{1, 2}$}
\email{daniel@stanciu.nl}

\author{A. Rebei$^1$}
\email{arebei@mailaps.org}

\affiliation{1. Seagate Research Center, Pittsburgh, PA 15222 \\
             2. Institute for Molecules and Materials,Radboud
             University Nijmegen, Toernooiveld 1, 6525 ED Nijmegen,
             The Netherlands}

\begin{abstract}
Magnetization reversal in GdFeCo by circularly polarized light is
shown to occur at the femtosecond time scale. In contrast to the
well known laser-assisted magnetization reversal based on the
laser heating, we here demonstrate that this femtosecond
all-optical magnetization reversal is more efficient at lower
temperatures. The lower the temperatures, the smaller the laser
fluence required for the switching. This switching is in agreement
with a more recent theoretical prediction $\left[ \text{Phys.
Lett. A \textbf{372}, 1915 (2008)} \right] $ and demonstrates the
feasibility of the femtosecond athermal magnetization reversal.
\end{abstract}
\date{\today}
\pacs{76.60.+q,78.20.Ls,75.40.Gb, 76.60.Es, 52.38.-r}
\maketitle

The current development in  electronic devices demands increasingly 
fast approaches to manipulate logical bits. In a
magnetic-memory device the speed of this process relies on fast
switching of the magnetization vector. Currently, the switching of
a magnetic bit in a hard disk drive occurs as quickly as $~500$\,ps, via the
application of an external magnetic field. A possible much faster
approach for the magnetization reversal has been indicated a
decade ago, with the demonstration of the laser induced ultrafast
demagnetization in magnetic metals \cite{Beaurepaire}. Yet, this
phenomena involved only the breakdown of the magnetization vector,
via heating, but not its control.

Employing $40$ femtosecond circularly polarized laser pulses it has
been recently demonstrated that laser alone can be used to control
the direction of the  magnetization \cite{stanciu}. It was shown that the
reversal of the magnetization can be controlled by the helicity of
the light and no external magnetic field is required. This new
magnetization reversal mechanism was understood as the combined
result of femtosecond laser \textit{heating} of the magnetic
system to just below the Curie point and circularly polarized
light simultaneously acting as a magnetic field. On the other
hand, it has been argued recently that this magnetization reversal
is completely \textit{athermal}, and therefore does not require
heating near Curie temperature \cite{rebei1}. More specifically,
it was proposed that for suitable parameters of the coupling
between the \textit{d} spins and the \textit{f} spins of the
rare-earth, and in the presence of a strong laser-enhanced
spin-orbit coupling of \textit{d} electrons (such as recently
demonstrated in ferromagnetic Ni \cite{Stamm}), femtosecond
athermal switching is possible in rare-earth doped transition
metals, via an inverse Einstein-De Haas effect (i.e. Barnett
effect \cite{Barnett}). A microscopic theory on the laser-induced
magnetization in a metallic material, to some extend similar to
the Barnett effect, has been also recently proposed \cite{Hertel}.
Nevertheless, the mechanisms responsible for the all-optical
\textit{permanent} magnetization reversal in metals are currently
a matter of debate. Besides this, another open question is related
to the all-optical switching reversal speed. In Ref.
\cite{stanciu} it is shown that each circularly polarized 40 fs
laser pulse leads to the formation of a permanent magnetic domain.
But what is the timescale of this reversal?

In this letter, using a time resolved pump-probe set-up, we
investigate the reversal time of all-optical switching. It is
experimentally demonstrated that the reversal occurs in the
sub-picosecond regime. In agreement with a very recent observation
\cite{Stamm}, this sub-picosecond switching rate indicates that
relaxation effects induced by optical excitations are much larger
than those derived from ferromagnetic resonance measurements. In
addition, by investigating the temperature dependence of the
all-optical switching we show that this process is taking place
even at low temperatures. More specifically, the lower the
temperatures, the smaller the laser fluence required for the
switching. This observation demonstrates the pure athermal origin
of the reversal mechanism.

The  time resolved experiments where performed in Pittsburgh,
employing relatively long laser pulses, of  $\approx500$~fs
duration, while the temperature dependent measurements were
carried out in Nijmegen using $40$~fs laser pulses. The metallic
magnet studied here was an amorphous GdFeCo ferrimagnetic alloy,
with the same composition as in Ref. \cite{stanciu}.

\begin{figure}[th]
   \includegraphics[width=.45 \textwidth]{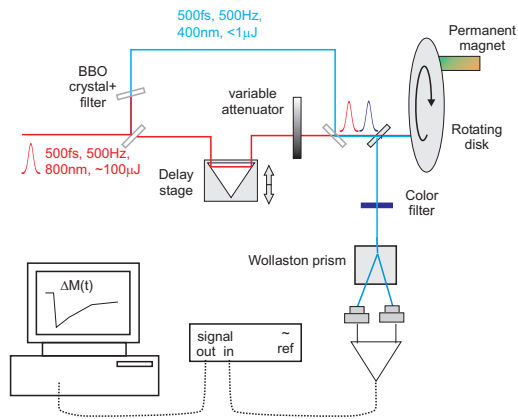}\\
    \caption{Schematic drawing of the time-resolved pump-probe experimental
    set-up used to measure the dynamics of all-optical switching on the femtosecond time scale.
    } \label{fig1}
\end{figure}

We first start by discussing the time resolved measurements. The
corresponding set up is shown in Fig. \ref{fig1}. The pump-probe
pulses use a two-color scheme with a wavelength of $\lambda =
800\,$~nm pump-, and $\lambda = 400\,$~nm probe-pulses. The
pump-pulses of $~100\,\mu$J energy are focused at normal incidence
onto the sample, via a $700\,$~mm lens. The collinear probe-pulses
are attenuated to $1\,\mu$J to avoid significant self-action, and
are focused with the help of a $100\,$~mm lens. The diameter of
the focused pump- and probe-beams was estimated to be of the order
of $~500\,\mu$m and $~50\,\mu$m, respectively. The Kerr-rotation
of the probe-pulses was recorded by a standard two diode scheme
\cite{stanciu2} and lock-in detection using the $500\,$~Hz
repetition rate of the laser system as a reference frequency. The sample (disk) is mounted on a 
fast rotation stage which
 ensures that every pump-probe pulse-pair is
exciting/probing a new spot initialized by a permanent magnet
mounted near the sample (see Fig. \ref{fig1}).

\begin{figure}[th]
    \includegraphics[width=.45 \textwidth]{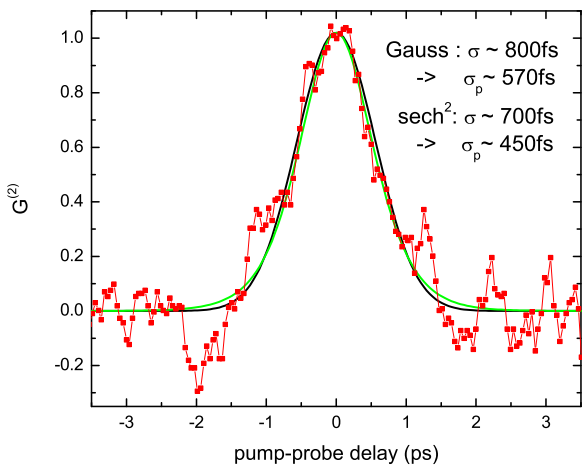}\\
    \caption{Second order intensity correlation, $G^{(2)}$, as
    measured for the laser pulse
    used in the real-time measurements (filled squares). The
    continuous lines  represent best fits to Gaussian and  sech$^2$
    pulse shapes which both give similar pulse widths.
    } \label{fig2}
\end{figure}

The pump pulse duration was determined by comparing measured and
calculated second-order intensity correlation functions as shown
in Fig. \ref{fig2}. The best fits for pulse shapes of the form of
a Gaussian and of a sech$^2$ lead to actual pulse lengths of
$570\,$fs, and $450\,$fs, respectively. Hence, we assume the
pulses to be of $\approx 500\,$fs duration, one order of magnitude
larger than the Nijmegen laser pulse width.

\begin{figure}[th]
    \includegraphics[width=.45 \textwidth]{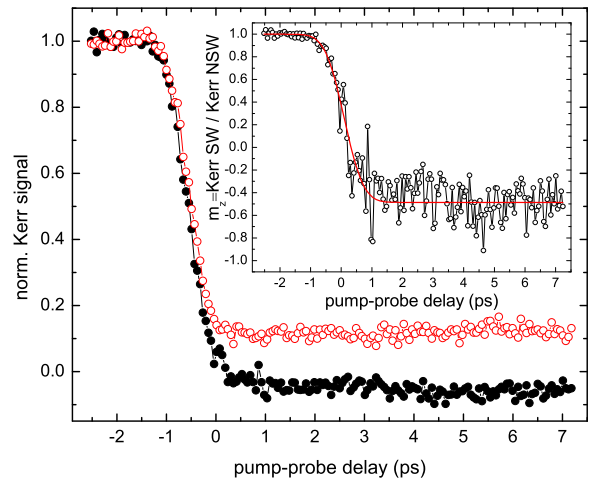}\\
    \caption{Transient magnetization dynamics induced by circularly polarized pump-pulses with
    helicity favoring the initial orientation of the magnetization
    (NSW, open symbols) and
    leading to magnetization reversal (SW, closed symbols). The inset shows the resulting
    dynamics of the normalized out-of plane component of the magnetization, $m_z$,
    obtained by dividing the transient for the switching case by the one for the non-switching
    configuration. The line is to guide the eye and represents an error-function of
    $700\,$fs width.
    } \label{fig3}
\end{figure}

Using relatively long pump-pulses, the optically
excited electrons have lots of time to thermalize among each other and
even to transfer a significant amount of energy to the lattice
while the pump-pulse is on. Hence, we observe a significant
reduction of the magnetization magnitude during the reversal
process (see Fig. \ref{fig3}). This apparent disadvantage is more
than compensated by the successful demonstration of all-optical
switching with $\approx 500\,$fs pulses that are far closer to
real applications than the more exotic $40\,$fs pulses used
previously \cite{stanciu}.

In order to separate switching dynamics from purely temperature
induced changes in the magnitude of the magnetization, we use a
procedure that was successfully applied to analyze the dynamics of
magneto-optical recording \cite{julius}. Here, we use the
normalized out-of plane component of the magnetization, $m_z$,
derived from the ratio of the  Kerr signals measured for both pump
helicities, i.e. for the switching and non-switching cases, to
extract the switching dynamics. It is clear from the time
dependence of $m_z$, shown in  the inset of Fig.~\ref{fig3},  that
the switching is very fast and occurs within one picosecond. For
our laser pulse of width $500$ fs, we estimate that the switching
is complete within $700$ fs.  The switching is therefore too fast
to follow a precessional path. Indeed, the model discussed in Ref.
\cite{rebei1} shows that precessional dynamics is not important
for femtosecond magnetization reversal. Figure~\ref{fig3} also
shows that our measurements yield a minimum value of $m_z=-0.5$
instead of $-1.0$, thereby indicating that on average only $75\%$
of the pump pulses lead to a reversed state. Because the
all-optical switching is strongly sensitive to a change in the
pump fluence \cite{stanciu}, a complete $100\%$ reversal requires
a very fine tuning of the laser fluence, which was difficult to
attain in our experiments. Shorter laser pulses should
yield stable switching over  a wider fluence range   and be able 
to diminish the effect of heat on the magnetization.

To explain this femtosecond magnetization reversal by circularly
polarized light, three main requirements must be fulfilled: 1) an
ultrafast channel for angular momentum exchange, between the spins
and another degree of freedom, such as lattice; 2) a light induced
switching mechanism, where the magnetization direction is
controlled by the light helicity; 3) a mechanism that, after
switching, maintains information about light helicity in spite of
the decoherence effects that takes place in metallic magnets on
the femtosecond time scale. We will now briefly discuss each of
these mechanisms:


1) We here observe an all-optical magnetization reversal taking
place on the femtosecond time scale. Therefore, it implies an
ultrafast transfer of angular momentum from the spin system to
another degree of freedom such as the orbital momentum of
electrons or lattice. In turn this also implies a strong
spin-orbit interaction during the optical excitation. Such fast
coupling is in agreement with the recent experimental
demonstration of a laser enhanced spin-orbit coupling in Ni
\cite{Stamm}. Hence we believe, as also discussed in Ref.
\cite{rebei1}, that relaxation of the non-equilibrium \textit{d}
electrons plays an important role in increasing the number of
channels available to relax the magnetization through spin-orbit
coupling or momentum relaxation \cite{rebei2}. A sub-picosecond
switching rate also indicates that relaxation effects induced by
optical excitations are much larger than those derived from
ferromagnetic resonance measurements (FMR). This is in line with the different energy 
scales of $1$\,eV and $1$\,meV for optical switching and FMR, respectively.

2) How is it possible that the angular momentum of the photons can
efficiently change magnetization? The optical electric dipole
transition can not affect the electronic spin. Magnetic dipole
transition may affect the spin but it requires annihilation of the
photon.   However, there are
not enough photons in the laser pulse to provide enough angular momentum
for a magnetization reversal \cite{Koopmans1}. An efficient switching
mechanism may take place via a stimulated Raman-like scattering
process \cite{stanciu2}. Yet, this switching mechanism requires
heating of the spin system to temperatures close to Curie
temperature \cite{stanciu}. As we will further demonstrate, the
all-optical switching takes place at a temperature of $~$200~K
lower than that used in Ref. \cite{stanciu}. Based on this result,
a more realistic scenario may be accounted to an optical Barnett
effect which works best at zero temeprature \cite{rebei1}.

3) Another issue that needs to be clarified is how the femtosecond
helicity-induced coherence among the itinerant electrons may be
converted to a static magnetization. It is well known that the
lifetime of a state at 1~eV above the Fermi level is a few
femtoseconds \cite{Aeschlimann}. Relaxation of these excited
states leads to the destruction of the spin coherence via
inelastic scattering processes. Indeed, it is by now well accepted
that there is no helicity-induced magnetization reversal in pure
transition-metals such as Ni, under similar conditions
\cite{francesco}. Thus, the observed memory effect in our
experiments may be explained as follows: During the optical
excitation, the negative exchange between the excited \textit{d}
electrons of both Gd and Fe collapses \cite{Durr}. On the other
hand, the rare-earth 4\textit{f} bands are about 4 eV below the
Fermi level and are therefore not directly excited by the
$~$1.5~eV photon energy used in our experiments. Yet, a strong
hybridization between the \textit{d}-spins and the
\textit{f}-spins of the rare-earth causes the component of the
\textit{f}-spins along the chirality of the laser to change sign
at high enough laser powers. Therefore, although the spin
coherence of the excited states may be lost via relaxation in
electron-electron scattering, \textit{f}-spins will maintain
sufficient coherence in the reversal process and therefore
maintain information about the chirality of the laser. In this
way, \textit{f}-spins will serve as nucleation points for what
will later become, via proceses such as domain wall propagation, a
complete reversed magnetic domain. This explains the importance of
the rare earth moments. A convincing test of this model will be
the investigation of the all-optical switching in SmCo$_5$ and
YCo$_5$. The
former has uncompensated \textit{f}-spins, but the latter
does not while both have similar crystal structure and
magneto-crystalline anisotropy.

As discussed above, an important factor indicating whether the
reversal mechanism takes place via a stimulated Raman-like
scattering or via an optical Barnet effect, is the relevance of
temperature in the reversal process. We therefore turn our
attention towards the low temperature all-optical switching
experiments.

\vskip 0.5cm

\begin{figure}[th]
    \includegraphics[width=.4 \textwidth]{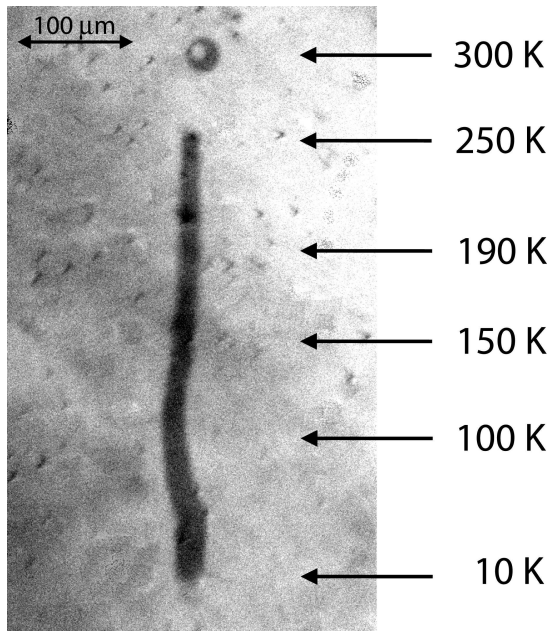}\\
    \caption{All-optical switching of magnetization as a function of temperature, in GdFeCo.
    Black and white areas correspond to oppositely oriented
    magnetic domains, perpendicular to the sample plain. The
    dot-like magnetic domain was created in order to indicated the laser
    beam location onto the sample, at 300~K. The switching behavior
    as a function of temperature has been investigated at a laser
    fluence of $2.5\,$ mJ/cm$^2$. At this laser fluence no
    switching is observed at room temperature but it
    becomes visible only when the temperature is reduced below 250~K.} \label{fig4}
\end{figure}

The temperature dependence of the all-optical switching has been
performed in a Gd$_{22}$Fe$_{74.6}$Co$_{3.4}$ film with a $20\,$nm
thickness. The sample structure was described in Ref.
\cite{stanciu}. The sample has been placed into a cryostat and
exposed to $40\,$fs circularly polarized laser pulses generated at
a repetition rate of 1 kHz. The laser wavelength was $800\,$nm and
the laser helicity has been chosen to be that relevant for the
switching. Initially, at $300\,$ K, in order to indicate the
location of the laser spot onto the sample, GdFeCo has been
exposed to a laser fluence of about $3\,$ mJ/cm$^2$, for a few
seconds. Next, the laser fluence has been reduced to about $2.5\,$
mJ/cm$^2$, that is below the fluence threshold required for the
all-optical switching in this sample \cite{stanciu}. At this laser
fluence no switching could be observed. Under these conditions the
temperature on the sample was reduced down to $10\,$ K at a rate
of about 5 K per minute while the sample was slowly vertically
shifted (from down to up). One can observe in Fig. \ref{fig4},
that while from $300\,$K down to $250\,$K the laser fluence of
$2.5$\,mJ/cm$^2$ does not change the magnetic state of the sample,
the conditions change drastically below $250\,$K. More
specifically, below this temperature the switching spot appears
and increasingly broadens as the temperature is reduced down to
$10\,$K. This observation gives clear evidence that thermal
fluctuations introduce decoherence in the system that upsets the
reversal process. Hence increasing the temperature decreases the
efficiency of the all-optical switching.

It has been initially argued that besides the non-thermal effect
where the light acts as an axial magnetic field, the heat from the
thermal bath is also required to elevate the temperature of the
magnetic system near Curie temperature \cite{stanciu}. The authors
in \cite{rebei1} argued and qualitatively showed that switching is
possible even at zero temperature if we take into account the
non-equilibrium character of the laser-induced process.  The
temperature dependent data presented here clearly favor the latter
suggestion. Note that the magnetization switching path at zero
temperature is very similar to switching by a strong field at the
Curie temperature \cite{chantrell} but the physics in both cases is very 
different. The Barnett effect discussed in \cite{rebei1} induces a frequency 
dependent interaction between the rare-earth ions and the itinerant \textit{d} electrons
of the transition metal ions that, depending on the helicity of the light, gives rise to
 either a ferromagnetic or 
anti-ferromagnetic coupling.  

In conlcusion, we have demonstrated that all optical switching in
GdFeCo is not a thermally assisted process but in fact thermal
fluctuations degrade the efficiency of the reversal. Moreover, we
showed that the switching occurs on a sub-picosecond time scale
which can not be explained in terms of the much slower
precessional reversal. Both findings, i.e. fast and athermal
mechanism, make all-optical recording on rare-earth transition
metal alloys a promising technique for magnetic storage with high
data rates. Furthermore, the all-optical switching using the long
$~$500fs laser pulses shown here, together with the recent
successful demonstration of miniature plasmonic wave-plates
\cite{Ebbesen}, push the all-optical switching close to
applications.

One of the authors (C. D. S.) expresses his gratitude to Prof. Rasing for mentorship. We also thank Prof.  
A. Tsukamoto for preapring the samples used in this study and Dr. Paul Jones for assistance with the laser set up. A.R
 acknowledge fruitful discussions with Prof. Roy Chantrell regarding his temperature calculations in similar problems.

\bigskip
\end{document}